\documentclass[12pt]{article}

\usepackage{amssymb}

\let\a=\alpha   \let\b=\beta   \let\g=\gamma   
         \let\q=\theta
        
           \let\p=\pi

\newcommand{\Zint}{\mathbb{Z}}

\newcommand{\be}{\begin{equation}}
\newcommand{\ee}{\end{equation}}
\newcommand{\bea}{\begin{eqnarray}}
\newcommand{\eea}{\end{eqnarray}}
\newcommand{\ba}{\begin{array}}
\newcommand{\ea}{\end{array}}

\newcommand{\drawsquare}[2]{\hbox{%
\rule{#2pt}{#1pt}\hskip-#2pt
\rule{#1pt}{#2pt}\hskip-#1pt
\rule[#1pt]{#1pt}{#2pt}}\rule[#1pt]{#2pt}{#2pt}\hskip-#2pt
\rule{#2pt}{#1pt}}
\newcommand{\Yasymm}{\raisebox{-3.5pt}{\drawsquare{6.5}{0.4}}\hskip-6.9pt%
        \raisebox{3pt}{\drawsquare{6.5}{0.4}}}

\newcommand{\YasymmS}{\raisebox{-1.9pt}{\drawsquare{4}{0.4}}\hskip-4.4pt%
        \raisebox{2.1pt}{\drawsquare{4}{0.4}}}
\newcommand{\bYasymmS}{\overline{\YasymmS}}
\usepackage{graphics}
\begin{document}
\begin{titlepage}
\rightline{{hep-th/0310277}}\vskip 2cm
\centerline{{\large\bf A Class of Non-Supersymmetric Open String Vacua}}
\vskip 0.3cm
\vskip 1cm
\centerline{
P. Anastasopoulos\footnote{panasta@physics.uoc.gr}$^{,\a,\b}$,
A. B. Hammou\footnote{amine@physics.uoc.gr}$^{,\a}$  and
N. Irges\footnote{irges@physics.uoc.gr}$^{,\a}$}
\vskip 1cm
\centerline{$^\a$ Department of Physics, University of Crete,}
\centerline{71003 Heraklion, GREECE.}
\vskip  .5cm
\smallskip
\centerline{$^\b$ Laboratoire de Physique Th\'eorique Ecole
Polytechnique,}
\centerline{91128 Palaiseau, FRANCE.}
\vskip  1cm
\begin{abstract}
We analyze non-supersymmetric four dimensional open string models
of type IIB string theory compactified on $T^2\times K3$ with
Scherk-Schwarz deformation acting on an $S^1$ of the $T^2$ torus.
We find that there are always two solutions to the tadpole
conditions that are shown to be connected via Wilson lines in an
non-trivial way. These models although non-supersymmetric, are
free of R-R and NS-NS tadpoles.

\end{abstract}
\end{titlepage}


\section{Introduction}

One of the outstanding problems in string theory is the
construction of realistic non-supersymmetric string vacua. In
particular, for open strings, an important program is the
cancellation of tadpoles that appear in the massless limit of the
transverse (or tree) channel amplitudes associated to Klein-Bottle
(${\tilde {\cal K}}$), Annulus (${\tilde {\cal A}}$) and M\"obius
strip (${\tilde {\cal M}}$) world-sheets, corresponding to the
exchange of a closed string between two crosscaps, two boundaries
and a crosscap and a boundary respectively. This cancellation is a
necessary condition for the consistency and the stability of the
vacuum.

In supersymmetric unoriented closed and open string models NS-NS
and R-R tadpoles are equal by supersymmetry and consequently if
the R-R tadpoles cancel, so do the NS-NS tadpoles. In models where
supersymmetry is broken in the open string sector, one can argue
that NS-NS tadpole cancellation implies that supersymmetry should
also be broken in the closed string sector \cite{Kir2000, AA}.
However, when supersymmetry is broken already at tree level in the
closed string sector, the situation is more involved and one has
in general non-zero NS-NS tadpoles even if R-R tadpoles cancel.
Recently, many non-supersymmetric open string vacua have been
constructed without R-R tadpoles \cite{Antoniadis:1998ki, BST,
Scrucca:2001ni,Ib1, Blum1}. Less is known about vacua which, in
addition, have zero NS-NS tadpoles.

The massless spectrum of open string models can be computed either
by looking directly at the action of the orientifold group on the
massless excitations in the closed and open string sectors
\cite{Gimon:1996rq, Gimon:1996ay} or by performing appropriate
modular transformations on ${\tilde {\cal K}}$, ${\tilde {\cal
A}}$ and ${\tilde {\cal M}}$ to obtain the corresponding direct
(or loop) channel amplitudes ${{\cal K}}$, ${{\cal A}}$, ${{\cal
M}}$ and taking their massless limit \cite{Pradisi:1988xd}. In the
former approach the action of the $i$th element of the orientifold
group $g_i$ on a $Dp$-brane is encoded in matrices acting on its
Chan-Paton factors, which we call $\g_{g_i,p}$. In the latter
approach, the Torus ${\cal T}$ and ${{\cal K}}$ contain the
information about the closed string spectrum and ${{\cal A}}$ with
${{\cal M}}$ contain the information about the open string
spectrum \cite{Pradisi:1988xd, Angelantonj:2002ct}.

The Scherk-Schwarz (SS) deformation \cite{Scherk:1979zr} is so far
the most interesting mechanism for supersymmetry breaking in which
supersymmetry is broken by twisting the boundary condition of the
fermions along some compact direction. In a recent paper
\cite{BST} the quantum stability of models with SS supersymmetry
breaking have been considered. It has been argued that the one
loop cosmological constant has a term power-like in the
compactification radii proportional to the difference between
fermionic and bosonic degrees of freedom and an exponentially
suppressed term. In \cite{AA} examples of non-supersymmetric but
fermion-boson degenerate models has been presented for the case of
M-theory breaking. In the class of models we consider in this
paper the massless spectrum we find is non-degenerate which would
imply that we will have a radius dependent one-loop cosmological
constant. We think that this question deserves more investigation.

In this letter we present a class of models in which supersymmetry
is broken by a Scherk-Schwarz deformation \cite{Scherk:1979zr} and
have zero tadpoles. In section 2, we discuss a nine dimensional
model, the simplest possible example in which the main points can
be illustrated. In section 3 we present a novel class of five
dimensional models and in section 4 we state our conclusions.

\section{Symmetry breaking in nine dimensions}

Consider the orientifold of the $S^1/Z^{\prime}_2$
compactification of type IIB string theory
\cite{Antoniadis:1998ki}, where $Z^{\prime}_2$ is the
freely-acting orbifold generated by an element $h$ acting as a
translation of length $\pi R$ along $S^1$, together with $(-1)^F$,
where $F$ is the space-time fermion number
\cite{Angelantonj:2002ct}. This orbifold, known as Scherk-Schwarz
deformation \cite{Scherk:1979zr, Kiritsis:1997ca}, breaks
spontaneously supersymmetry by assigning different boundary
conditions to bosons and fermions.

The loop channel Klein-Bottle amplitude is obtained by projecting
the torus amplitude
by $\Omega$ (the notation we follow in this section is the one
used in ref. \cite{Angelantonj:2002ct}):
\be {{\cal K}}\sim {1\over 4}(V_8-S_8) P_{2m}~. \label{K} \ee
Here $V_8$ and $S_8$ are the standard bosonic and fermionic
$SO(8)$ characters respectively and $P_{2m}$ is the momentum
lattice with even momenta. By a modular transformation one obtains
the tree channel Klein-Bottle amplitude
\be {\tilde {\cal K}}\sim \frac{2^5}{4} R (V_8-S_8) W_n~,
\label{tildeK} \ee
where $W_n$ is the winding lattice. The above amplitude contains
massless R-R tadpoles and corresponds to an $O9$-plane with
positive tension and charge, i.e. to an ${\it O^+}$-plane. To
cancel this tadpole, a stack of 32 $D9$-branes has to be
introduced. The most general Annulus amplitude associated with
these $D9$-branes including Wilson lines is \footnote{This is
actually the most general Wilson line that in the T-dual model
moves the stack around the T-dual circle as a whole.}
\bea {\cal A} &\sim &\frac{1}{4}~\bigl[ (N^2 P_{m-2\q}+ \bar{N}^2
P_{m+2\q}+ 2N\bar{N} P_m)
(V_8-S_8) \nonumber\\
&&+(N^2 P_{m-2\q}+ \bar{N}^2 P_{m+2\q}+ 2N\bar{N}
P_m)(-1)^m(V_8+S_8)~\bigr]~ . \label{tildeA} \eea
The transverse channel amplitude is obtained by a modular
transformation, yielding
\bea {\tilde {\cal A}}& \sim &\frac{2^{-5}}{4} R
~\bigl[ (N e^{2\pi in{\q}}+ \bar{N} e^{-2\pi in{\q}})^2(V_8-S_8)W_n\nonumber\\
&&+(N e^{2\pi in{\q}}+ \bar{N} e^{-2\pi in{\q}})^2(O_8-C_8) W_{n+{1\over
2}}~\bigr]~. \label{annulus} \eea
The tree channel M\"obius strip amplitude is then obtained as a
state by state geometric mean of the Klein-Bottle amplitude
${\tilde {\cal K}}$ and the tree channel Annulus amplitude
${\tilde {\cal A}}$:
\be {\tilde {\cal M}}\sim -\frac{2}{4} R (N e^{2\pi in{\q}}+ \bar{N}
e^{-2\pi in{\q}}) ({\hat V}_8 - (-1)^n {\hat S}_8) W_n~. \label{tildeM} \ee
The $(-1)^n$ is introduced due to a sign ambiguity in taking the
mean value\footnote{Ignoring this sign will generate a
supersymmetric M\"obius strip amplitude. Note also that it seems
to be possible to put the sign in front of $V_8$ instead. However,
it turns out that this choice is not consistent with the
parametrization we have chosen in the Annulus amplitude.}.
Performing a modular transformation, one finds the direct channel
amplitude
\be {\cal M}\sim -\frac{1}{2} ~\bigl[(N P_{2m-2\q}+ \bar{N}
P_{2m+2\q}){\hat V}_8 - (N P_{2m-2\q+1}+ \bar{N} P_{2m+2\q+1}) {\hat S}_8~\bigr]~.
\label{M} \ee
The Wilson line takes values in $[0,1]$ mod $\Zint$. We will
distinguish two cases corresponding to $\q=0, {1\over 2}$. The
first case ($\q =0$) gives $SO(N+\bar{N})$ gauge group with
$N=\bar{N}=16$ and no massless fermions. The second case
($\q={1\over 2}$) gives a $U(16)$ gauge group with fermions in the
symmetric representation. This is because the $Z'_2$ projection
gives antiperiodic boundary conditions to the fermions but its
effect is cancelled by the $\q={1 \over 2}$ Wilson
line\footnote{We would like to thank Carlo Angelantonj for very
helpful discussion on this point.}. Note that in the
supersymmetric case $\q={1\over 2}$ leads to $SO(32)$. This
mismatch in the values of $\q$ is due to the shift, since putting
a shift results in an effective rescaling of the radius by a
factor of 2. It is easy to see that for both cases tadpoles
cancel.

Before ending this section, let us make connection with the
Chan-Paton algebra formalism \cite{Gimon:1996rq}. With vanishing
Wilson lines, besides the usual untwisted tadpole condition that
fixes the number of $D9$-branes to $Tr[\g_{1,9}]=32$, one finds
from the M\"obius strip amplitude the constraints
\bea
\g^{T}_{\Omega,9}&=&\g_{\Omega,9}~ \nonumber\\
\g^{T}_{\Omega h,9}&=& \pm \g_{\Omega h,9}~, \label{htadpoles}\eea
which imply that $\g^{2}_{h,9}=\pm 1$ \cite{BST}, thus giving two
possible choices for the $\g_{h,9}$ matrix. Note that tadpole
cancellation does not impose any constraint on $Tr[\g_{h,9}]$. A
solution to equations (\ref{htadpoles}) is
\bea
\g_{h,9}^2=+{\bf 1}_{32}: \hskip .5cm \g_{h,9}&=& \textrm{diag}(-{\bf 1}_n,\;
{\bf 1}_{32-n})
\label{TadpoleConditions1}\\
\g_{h,9}^2=-{\bf 1}_{32}: \hskip .5cm \g_{h,9}&=& \textrm{diag}
(e^{{i\pi}\over 2} {\bf 1}_{16},\; e^{-{i\p\over 2}} { \bf
1}_{16}) \label{TadpoleConditions2}\eea
where ${\bf 1}_n$ the $n\times n$ identity matrix with $n$ an even
integer. Solution (\ref{TadpoleConditions1}) for $n=0$ and
solution (\ref{TadpoleConditions2}) lead to the two distinct gauge
groups and spectra we found earlier in our simple model
corresponding to integer and half integer $\q$ respectively.

For general $n$ \footnote{$n\ne 0$ amounts to splitting the stack
of $D9$-branes into two smaller stacks.}, the two solutions are
just particular realizations of the two possible breaking patterns
of an even dimensional orthogonal group projected out by a $Z_2$
inner automorphism \cite{Slansky:yr}. We could have easily found
all these solutions in the simple model as well by choosing an
appropriately more general Wilson line in (\ref{tildeA}). The
conclusion therefore is that the seemingly two independent
solutions (\ref{TadpoleConditions1}) and
(\ref{TadpoleConditions2}) are in fact related via Wilson lines.
Nevertheless, they define two classes of physically inequivalent
massless spectra and thus they are both interesting in their own
right.

\section{Non-supersymmetric $T^2\times K3$}

Consider the ${\cal N}=4$ orbifold of type IIB string theory in
four dimensions, $R^4 \times T^2\times (T^4 /Z_N)$. The $Z_N$
orbifold acts on the complex coordinates $z^1=x^6+ix^7$ and
$z^2=x^8+ix^9$ of the $T^4$ torus as $\theta^k: z^i\to  e^{2\pi
ikv_i} z^i$, where $v={1\over N}(1,-1)$ and $k=1,\cdots ,N-1$
labels the different $Z_N$ orbifold sectors. We will concentrate
on orbifolds with $N=2,3,4,6$. In addition, we act with a
freely-acting $Z'_2$ orbifold generated by the SS element $h$
acting as a translation of length $\pi R$ along the direction
$x^5$ of $S^1$ in the $T^2$ torus together with a $(-1)^F$. We
shall consider in the following an orientifold of the type
$G+\Omega G$, where $G$ is $Z_N\times {Z^\prime}_2$ which breaks
supersymmetry completely.

Upon projecting this orbifold by the world sheet parity $\Omega$,
the massless limit of the tree channel Klein Bottle amplitude has
non-vanishing R-R tadpoles and thus reveals the presence of
orientifold planes in the background. Besides the $O9$-plane that
extends in the non-compact directions, wraps the $T^2\times T^4$
and it is present for any $N$, for even $N$ the model contains
also $O5$-planes that extend along the non-compact directions,
wrap around the $T^2$ and sit at the $\theta^k$-fixed points of
the transverse $T^4$. In order to cancel the associated to the
orientifold planes massless tadpoles one has to introduce $D9$ and
$D5$-branes. The contribution of the $D$-branes to the tadpoles is
encoded in the massless limit of the transverse channel Annulus
and M\"obius strip amplitudes.

For sake of brevity we will skip the details of the calculation
and present directly the result for the massless tadpole
conditions. The action of the $Z_N\times Z^{\prime}_2$ orbifold
$g_i=(1 ,\theta^k ,h ,\theta^k h)$ on the Chan-Paton matrices
carried by the $D9$ and $D5$-branes is described by $32\times 32$
matrices $\g_{g_i,9}$ and $\g_{g_i,5}$.
%
%
The matrices $\g_{1,9}$ and $\g_{1,5}$ that correspond to the
identity element of $Z_N\times Z'_2$ can be chosen to be the
$32\times 32$ identity matrices, so that
$Tr[\g_{1,9}]=Tr[\g_{1,5}]=32$. This is a constraint on the number
of D-branes that originates from tadpole cancellation in the
untwisted sector. The twisted tadpole conditions on the other hand
in the $\theta^k$ twisted sector, for $N$ even are given by
\cite{Gimon:1996ay}
\bea &&Tr[\g_{\theta^{2k-1},9}]-4\sin^2 {(2k-1)\pi \over N}~
Tr[\g_{\theta^{2k-1},5}]=0
\label{tad1}\\
&&Tr[\g_{\theta^{2k},9}]-4\sin^2 {2\pi k\over N}~
Tr[\g_{\theta^{2k},5}]-32\cos \frac{2\pi k}{N} =0,
\label{tad2}\eea
whereas for $N$ odd they read
\bea Tr[\g_{\theta^{2k},9}]-32\cos^2 \frac{\pi k}{N}=0.
\label{tad3}\eea
From the $\theta^k h$ and $h$ twisted sectors we do not get
further constraints on $Tr[\g_{\theta^{k} h,9}]$,
$Tr[\g_{\theta^{k}h,5}]$, $Tr[\g_{h,9}]$ and $Tr[\g_{h,5}]$.
Notice that for N even, the tadpole conditions are consistent with
T-duality transformations along the $T^4$ torus that exchanges the
$D9$ and $D5$-branes. On the other hand, for the circle along
which the shift is performed, we have a freedom in taking
$\g^{2}_{h,9}=\pm 1$ and also $\g^{2}_{h,5}=\pm 1$, however
T-duality constrains them to have the same sign. In summary, we
will obtain two open string spectra for each $N$, related by
Wilson lines, as we have explained in the previous section.

Let us describe the massless spectrum starting from the closed
string sector. The closed string spectra of the supersymmetric
$T^4/Z_N$ orientifolds have been computed in \cite{Gimon:1996rq,
Gimon:1996ay}. Sectors twisted by $h$ do not contribute to the
massless part of the Torus and the Klein-Bottle since they
correspond to half integer winding \cite{Angelantonj:2002ct}.
Every other massless sector in the Torus is the same as in the
corresponding supersymmetric model \footnote{By corresponding
supersymmetric model we simply mean the model obtained by
eliminating the SS part, which is supersymmetric for all values of
$N$ discussed here.} plus an identical sector where the sign of
the fermions is reversed.
This simply means that $h$ projects out the fermions altogether
from the closed string sector. The bosons remain multiplied by a
factor of two which is cancelled by the 1/2 of the $h$-projector
$(1+h)/2$ in the trace. The Klein-Bottle on the other hand remains
the same as in the corresponding supersymmetric model. The extra
1/2 from the $h$-projector is now cancelled by a factor of two
coming from the doubling of the surviving the $\Omega$ projection
states, since any sector and its projected by $h$ counterpart give
the same contribution to the Klein-Bottle.
The closed string spectrum therefore for any $N$ is just the
bosonic part of the corresponding supersymmetric model compactified
on a $T^2$ torus.

The full open string spectrum will be presented in tables 1 and 2
for each value of $N$ considered here. As we mentioned before we
have two inequivalent spectra for each $N$ corresponding to
$\g^{2}_h=\pm 1$. The effect of the SS deformation on the open
strings in a given supersymmetric model is to break the gauge
group for $\g^{2}_h=+1$ as
\be U(N)\rightarrow U(n)\times U(N-n),\hskip 1cm SO(N)\rightarrow
SO(n)\times SO(N-n), \ee
whereas for $\g^{2}_h=-1$ as
\be U(N)\rightarrow U(n)\times U(N-n),\hskip 1cm SO(2N)\rightarrow
U(N).\ee
For example, for $N=2$ and $\g^{2}_h=+1$ the $99$ and $55$ sectors
contain gauge bosons and scalars (corresponding to the $T^2$
torus) in the adjoint of $U(a)\times U(b)$ with $a+b=16$ and the
remaining scalars (corresponding to the $T^4$ torus) in the
$(~\Yasymm,1)$ and $(1,\Yasymm ~)$ where $\Yasymm$ is the
antisymmetric representation of the corresponding gauge group,
together with their complex conjugates.
The fermions are in the bifundamental representation $(a,b)$ and
$2\times (a,\bar{b})$ plus their complex conjugates. The $95$
sector contains bosons in $(a,1;\bar{a},1)$ and $(1,b;1,\bar{b})$
and fermions in $(a,1;1,\bar{b})$ and $(1,b;\bar{a},1)$ plus their
complex conjugates. On the other hand, for $\g^{2}_h=-1$ the gauge
group is again $U(a)\times U(b)$ with $a+b=16$. All the scalars
are in the $(a,b)$ and the fermions are in the $(~\Yasymm,1)$,
$(1,\Yasymm ~)$ and $2\times(a,\bar{b})$ representations plus
their complex conjugates. The $95$ sector is identical to the
previous case. It is easy to check that the above spectrum as well
as the spectra for $N=3,4,6$ do not suffer from irreducible gauge
anomalies. This is due to the fact that all fermions are in vector
like representations. Alternatively, the models we have considered
are effectively five dimensional and therefore do not have
anomalies.

\section{Conclusion}

We have presented a class of non-supersymmetric open string vacua
without tadpoles. In particular, satisfying conditions
(\ref{tad1}-\ref{tad3}) implies the vanishing of the twisted R-R
and NS-NS tadpoles, even though supersymmetry is broken both in
the closed and the open string sectors. This should not come as a
surprise. In the closed string sector the SS deformation just
lifts the fermions and therefore it does not affect the R-R or
NS-NS states which are the ones that contribute to the tadpoles.
In the open string sector there are no ${\bar D}$-branes necessary
to cancel the orientifold plane charge which means that the tree
channel Annulus amplitude does not contain sectors projected by
$h$.
These sectors contain massless states and if they were present,
could alter the supersymmetric tadpole cancellation conditions. On
the other hand, the tree channel Annulus amplitude does have
sectors twisted by $h$, which however do not contain massless
states and so do not contribute to tadpoles. In fact, the SS
deformation does not seem to alter the tadpole cancellation
conditions for any model in which the SS acts along a direction
orthogonal to the space where $Z_N$ acts.

We showed that the spectrum for each $N$ splits into two
inequivalent branches. The existence of the two branches was
understood to have a group theoretic origin associated to the
different ways one can embed a $Z_2$ inner automorphism into the
$SO(2n)$ and $U(2n)$ Lie algebras and it was shown that the
associated vacua are related by Wilson lines.

It would be interesting to extend this analysis to $T^6/Z_N$ and
$T^6/Z_N\times Z_M$. In these cases the SS deformation will act in
the same direction as the orbifold group. The allowed orbifolds
are the ones that commute with the SS deformation
\cite{Benakli:1995ut}. Models where the SS deformation acts along
a $Z_2$ direction have been constructed in
\cite{Antoniadis:1998ki, Scrucca:2001ni}.

\vskip .5cm

\centerline{\bf\Large Acknowledgments}


The authors would like to thank Carlo Angelantonj and Elias
Kiritsis for very useful discussions. We also would like to thank
Adi Armoni for valuable comments. The work of A.B. Hammou was
supported by RTN contracts HPRN-CT-2000-0131. This work was
partially supported by RTN contracts HPRN-CT-2000-00122 and INTAS
contract 99-1-590.





%
%
%
%
%
\begin{table}[h]\footnotesize \renewcommand{\arraystretch}{.8}
\begin{tabular}{|c c c|}

\hline \hline
& & \\
& \raisebox{.8ex}[0cm][0cm]{\textbf{Z}$_3$} &  \\
\hline \hline \hline
$\g_h^2=-1$ & & \\
$U(a)\times U(b)\times U(8)$ &
\raisebox{.8ex}[0cm][0cm]{~~~~~~~~~~~~~~(99)/(55)~matter~~~~~~~~~~~~~~}
& ~~~~~~~~~~~~~~~~~~~~~~~~~~~~~~~~~~~~~~~~~~~~~~~~\\
\hline \hline
& & \\
\raisebox{.8ex}[0cm][0cm]{Scalars} & \raisebox{.8ex}[0cm][0cm]{
adjoint $+(a,b,1)+ (\bar{a},1,8)+ (1,b,\bar{8})+c.c.$} &
\\
\hline
& $2 \left( (a,\bar{b},1)+ (1,1,\YasymmS)\right)+(\YasymmS,1,1)+$
& \\
\raisebox{.8ex}[0cm][0cm]{Fermions} & $+(1,\YasymmS,1)+
(\bar{a},1,\bar{8})+ (1,\bar{b},8)+c.c.$ &
\\
\hline \hline
$\g_h^2=+1$ & & \\
$U(a)\times U(b)\times$ & (99) matter & \\
$SO(c)\times SO(d)$ &&\\
\hline \hline
& adjoint $+(\YasymmS,1,1,1)+ (\bar{a},1,c,1)$ & \\
\raisebox{.8ex}[0cm][0cm]{Scalars} & $+(1,\YasymmS,1,1)+
(1,\bar{b},1,d)+c.c.$ &
\\
\hline
& $2 \left((a,\bar{b},1,1)+ (1,1,c,d)\right)
+(\bar{a},\bar{b},1,1)$
& \\
\raisebox{.8ex}[0cm][0cm]{Fermions} & $+(a,1,1,d)+ (1,b,c,1)+c.c.$
&
\\
\hline
%
%
%
%
%
%
%
%
%
\hline \hline
& & \\
& \raisebox{.8ex}[0cm][0cm]{\textbf{Z}$_4$} &  \\
\hline \hline \hline
$\g_h^2=-1$ & & \\
$\big\{U(a)\times U(b)\times$ &
~~~~~~~~~~~~~~(99)/(55)~matter~~~~~~~~~~~~~~
& ~~~~~~~~~~~~~~~(59) matter ~~~~~~~~~~~~~~~\\
$U(c)\times U(d)\big\}_{9,5}$ &&\\
\hline \hline
&  adjoint $+(\bar{a},\bar{b},1,1)+(a,1,\bar{c},1)$ &
$(a,1_3;\bar{a},1_3)+(1,b,1_2;1,\bar{b},1_2)+$ \\
\raisebox{.8ex}[0cm][0cm]{Scalars} & $+(1,b,1,\bar{d})+
(1,1,c,d)+c.c.$ &
$(1_2,c,1;1_2,\bar{c},1)+(1_3,d;1_3,\bar{d})+c.c.$
\\
\hline
& $2\times \left((a,\bar{b},1,1)+(1,1,c,\bar{d})\right)$ &
\\
Fermions &
$+(\YasymmS,1,1,1)+(\bar{a},1,1,\bar{d})+(1,\YasymmS,1,1)$ &
\raisebox{.8ex}[0cm][0cm]{
$(a,1_3;1,\bar{b},1_2)+(1,b,1_2;\bar{a},1_3)$+} \\
& $(1,\bar{b},c,1)+(1,1,\YasymmS,1)+(1,1,1,\YasymmS)+c.c.$ &
\raisebox{.8ex}[0cm][0cm]{$(1_2,c,1;1_3,\bar{d})+
(1_3,d;1_2,\bar{c},1)+c.c.$} \\
\hline \hline
$\g_h^2=+1$ & & \\
$\big\{U(a)\times U(b)\times$ &
(99)/(55) matter & (59) matter \\
$U(c)\times U(d)\big\}_{9,5}$ &&\\
\hline \hline
&  adjoint $+(\YasymmS,1_3)+(\bar{a},1,c,1)+(1,\YasymmS,1_2)$ &
$(a,1_3;\bar{a},1_3)+(1,b,1_2;1,\bar{b},1_2)+$ \\
\raisebox{.8ex}[0cm][0cm]{Scalars} & $+(1,\bar{b},1,d)+
(1_2,\bYasymmS,1)+ (1_3,\bYasymmS)+c.c.$ &
$(1_2,c,1;1_2,\bar{c},1)+ (1_3,d;1_3,\bar{d})+c.c.$
\\
\hline
& $2 \left((a,\bar{b},1,1)+ (1,1,c,\bar{d})\right)+
(\bar{a},\bar{b},1,1)$ &
$(a,1_3;1,\bar{b},1_2)+ (1,b,1_2;\bar{a},1_3)+$\\
\raisebox{.8ex}[0cm][0cm]{Fermions} & $+(a,1,1,\bar{d})+
(1,b,\bar{c},1)+ (1,1,c,d)+c.c.$ &
$(1_2,c,1;1_3,\bar{d})+ (1_3,d;1_2,\bar{c},1)+c.c.$ \\
\hline
\end{tabular}\caption{The $h$ action on the Chan-Paton charges
breaks the gauge group of the six-dimensional supersymmetric
orientifolds compactified on $K3$. For $Z_3$ and $Z_4$
$a+b=c+d=8$.}
\end{table}
%
%
%
%
%
\begin{table}[h]\footnotesize \renewcommand{\arraystretch}{.8}
\begin{tabular}{|c c c|}
\hline \hline
& & \\
& \raisebox{.8ex}[0cm][0cm]{\textbf{Z}$_6$} &  \\
\hline \hline \hline
$\g_h^2=-1$ & & \\
$\big\{U(a)\times U(b)\times$ & & \\
$ ~~U(c)\times U(d)\times $ & \raisebox{.8ex}[0cm][0cm]{(99)/(55)
matter} & \raisebox{.8ex}[0cm][0cm]{ (59) matter } \\
$~~~~~ U(e)\times U(f) \big\}_{9,5} $& & \\
\hline \hline
&  adjoint $+(\bar{a},\bar{b},1_4)+ (a,1,\bar{c},1_3)+$ &
$(a,1_5;\bar{a},1_5)+ (1,b,1_4;1,\bar{b},1_4)+$ \\
Scalars & $(1,b,1,\bar{d},1_2)+ (1_2,c,1,\bar{e},1)+$ &
$(1_2,c,1_3;1_2,\bar{c},1_3)+ (1_4,e,1;1_4,\bar{e},1)$
\\
& $(1_3,d,1,\bar{f})+ (1_4,e,f)+c.c.$ &
$(1_3,d,1_2;1_3,\bar{d},1_2)+ (1_5,f;1_5,\bar{f})+c.c.$ \\
\hline
& $2 \left( (a,\bar{b},1_4)+ (1_2,c,\bar{d},1_2)+
(1_4,e,\bar{f})\right)+$ & $(a,1_5;1,\bar{b},1_4)+
(1,b,1_4;\bar{a},1_5)+$\\
& $(\bar{a},1_2,\bar{d},1_2)+ (1,\bar{b},c,1_3)+
(1_2,\bar{c},1_2,f)$ &

$(1_2,c,1_3;1_3,\bar{d},1_2)+(1_4,e,1;1_5,\bar{f})$ \\
\raisebox{.8ex}[0cm][0cm]{Fermions}& $+(1,b,1_4;\bar{a},1_5)+
(1_3,\bar{d},e,1)+ (\YasymmS,1_5) $ &
$(1_3,d,1_2;1_2,\bar{c},1_3)+(1_5,f;1_4,\bar{e},1)$\\
& $(1,\YasymmS,1_4)+ (1_4,\bYasymmS,1)+ (1_5,\bYasymmS)+c.c.$ &
$+c.c.$
\\
\hline
%
%
%
%
%
\hline
$\g_h^2=+1$ & & \\
$\big\{U(a)\times U(b)\times$ & & \\
$ ~~U(c)\times U(d)\times $ & \raisebox{.8ex}[0cm][0cm]{(99)/(55)
matter} & \raisebox{.8ex}[0cm][0cm]{ (59) matter } \\
$~~~~~ U(e)\times U(f) \big\}_{9,5} $& & \\
\hline \hline
& adjoint $+(\bar{a},1,\bar{c},1_3)+ (1,\bar{b},1,d,1_2)$ &
$(a,1_5;\bar{a},1_5)+ (1,b,1_4;1,\bar{b},1_4)$ \\
Scalars & $(1_2,\bar{c},1,e,1)+ (1_3,\bar{d},1,f)+(\YasymmS,1_5)$
& $(1_2,c,1_3;1_2,\bar{c},1_3)+ (1_4,e,1;1_4,\bar{e},1) $
\\
& $+ (1,\YasymmS,1_4)+ (1_4,\bYasymmS,1)+ (1_5,\bYasymmS)$ &
$ (1_3,d,1_2;1_3,\bar{d},1_2) + (1_5,f;1_5,\bar{f})$ \\
\hline
& $2\times \left(
(a,\bar{b},1_4),~(1_2,c,\bar{d},1_2),~(1_4,e,\bar{f})\right)$ &
$(a,1_5;1,\bar{b},1_4)+ (1,b,1_4;\bar{a},1_5)$
\\
Fermions & $(\bar{a},\bar{b},1_4)+ (a,1_2,\bar{d},1_2)+
(1,b,\bar{c},1_3)$ &
$(1_2,c,1_3;1_3,\bar{d},1_2)+ (1_2,c,1;1_3,\bar{d})$ \\
& $(1_2,c,1_2,\bar{f})+ (1_3,d,\bar{e},1)+ (1_4,e,f)$ &
$(1_3,d,1_2;1_2,\bar{c},1_3)+ (1_3,d;1_2,\bar{c},1)$ \\
\hline \hline
\end{tabular}\caption{For $Z_6$ $2a+2b=c+d=2e+2f=8$.}
\end{table}



\end{document}